\documentclass[12pt]{article}
  \usepackage{amsfonts}
  \usepackage{amsmath}
\usepackage{amssymb}
\usepackage{amscd}
\usepackage[dvips]{graphicx}
\begin{document}
~~
\bigskip
\bigskip
\begin{center}
%\section*
{\Large {\bf{{{Generalized twist-deformed Rindler space-times}}}}}
\end{center}
\bigskip
\bigskip
\bigskip
\begin{center}
{{\large ${\rm {Marcin\;Daszkiewicz}}$ }}
\end{center}
\bigskip
\begin{center}
{ {{{Institute of Theoretical Physics\\ University of Wroc{\l}aw pl.
Maxa Borna 9, 50-206 Wroc{\l}aw, Poland\\ e-mail:
marcin@ift.uni.wroc.pl}}}}
\end{center}
\bigskip
\bigskip
\bigskip
\bigskip
\bigskip
\bigskip
\bigskip
\bigskip
\begin{abstract}
The (linearized) quantum Rindler space-times associated with
generalized  twist-deformed Minkowski spaces  are provided. The
corresponding corrections to the Hawking spectra linear in
deformation parameters  are derived.
\end{abstract}
\bigskip
\bigskip
\bigskip
\bigskip
\bigskip
\bigskip
\bigskip
\bigskip
\bigskip
 \eject
\section{{{Introduction}}}

Presently, it is well known that there exist the deep (and
extraordinary) relation between horizons of black hole and
thermodynamics. Already in early  1970s there was observed by
Bekenstein (see \cite{bekenstein}), that laws of black hole dynamics
(especially the second one) can be given  thermodynamical
interpretation, if one identifies entropy with the area of black
hole horizon and temperature  with its "surface gravity". Such an
observation has been confirmed by Hawking in his two seminal
articles \cite{hawking1}, \cite{hawking2}, in which, it was
predicted that a black hole should radiate with a temperature
\begin{equation}
T_{{\rm Black\;Hole}} = \frac{\hbar g}{2\pi kc}\;, \label{tempe}
\end{equation}
where $g$ denotes the gravitational acceleration at the surface of
the black hole, $k$ is Boltzman's constant, and $c$ is the speed of
light. Subsequently, it was shown separately by Davies \cite{devies}
and Unruh \cite{unruh}, that uniformly accelerated observer in
vacuum detects a radiation (a thermal field) with the same
temperature  as $T_{{\rm Black\;Hole}}$
\begin{equation}
T_{{\rm Vacuum}} = \frac{\hbar a}{2\pi kc}\;, \label{tempe1}
\end{equation}
but with inserted   acceleration   of the detector $a$. Formally,
such an observer "lives" in so-called Rindler space-time
\cite{rindler}, which can be obtained by the following
transformation from Minkowski space with coordinates
$(x_0,x_1,x_2,x_3)$\footnote{$c=1$.}
\begin{eqnarray}
x_0 &=& N(z_1)\, {\rm sinh} (az_0)\;,\label{trans1}\\
x_1 &=& N(z_1)\, {\rm cosh} (az_0)\;,\label{trans2}\\
x_2 &=& z_2\;,\label{trans3} \\
x_3 &=& z_3\;,\label{trans4}
\end{eqnarray}
where $N$ is a positive function of the coordinate. The Minkowski
metric $ds^2 = -dx_0^2 + \sum_{i=1}^{3}dx_i^2$ transforms to
\begin{equation}
ds^2 =-aN^2(z_1)dz_0^2 + (N')^2(z_1)dz_1^2 + dz_2^2 + dz_3^2\;.
\label{metric}
\end{equation}

Recently, in the papers \cite{krindler} and \cite{daszrindler},
there was proposed the noncommutative counterpart of Rindler space
 - so-called (linearized) $\kappa$-Rindler space and
 twist-deformed Rindler space-time, respectively. First
of them is associated with the well-known $\kappa$-deformed
Minkowski space \cite{kmin1}, \cite{kmin2}\footnote{$\kappa$ denotes
the mass-like parameter identified with Planck's mass.}, while the
second one with twisted canonical, Lie-algebraic and quadratic
quantum Minkowski space-time. Further, following the content of the
papers \cite{bekenstein}-\cite{unruh} (see also \cite{simpl1},
\cite{simpl2}), there have  been found corrections
 to the Hawking thermal spectrum linear in deformation parameter, which are detected by (noncommutative and
uniformly accelerated) $\kappa$- as well as twist-deformed Rindler
observers.

The suggestion to use noncommutative coordinates goes back to
Heisenberg and was firstly  formalized by Snyder in \cite{snyder}.
Recently, however, the interest in space-time noncommutativity is
growing rapidly. Such a situation  follows from many
phenomenological suggestions, which state that  relativistic
space-time symmetries should be modified (deformed) at Planck scale,
while  the classical Poincare invariance still remains  valid at
larger distances \cite{1a}-\cite{1d}. Besides, there have been found
 formal arguments, based mainly on Quantum Gravity \cite{2},
\cite{2a} and String Theory models  \cite{recent}, \cite{string1},
indicating that space-time at Planck-length should be
noncommutative, i.e. it should have a quantum nature.

At present, in accordance with the Hopf-algebraic classification of
all deformations of relativistic and nonrelativistic symmetries
\cite{clas1}, \cite{clas2}, one can distinguish three kinds of
quantum spaces. First of them corresponds to the well-known
canonical type of noncommutativity
\begin{equation}
[\;{\hat x}_{\mu},{\hat x}_{\nu}\;] =
i\theta_{\mu\nu}\;,%\;\;\;;\;\;\; \theta_{\mu\nu} = {\rm const}\;,
\label{wielkaslawia}
\end{equation}
with antisymmetric constant tensor $\theta^{\mu\nu}$. Its
relativistic and nonrelativistic Hopf-algebraic counterparts have
been proposed in  \cite{chi} and \cite{daszkiewicz} respectively.\\
The second kind of mentioned deformations  introduces the
Lie-algebraic type of space-time noncommutativity
\begin{equation}
[\;{\hat x}_{\mu},{\hat x}_{\nu}\;] = i\theta_{\mu\nu}^{\rho}{\hat
x}_{\rho}\;, \label{noncomm1}
\end{equation}
with particularly chosen coefficients $\theta_{\mu\nu}^{\rho}$ being
constants. The corresponding Poincare quantum groups have been
introduced in \cite{kappaP1}-\cite{lie2}, while the suitable Galilei
algebras  - in \cite{kappaG} and \cite{daszkiewicz}.\\
The last kind of quantum space, so-called quadratic type of
noncommutativity
\begin{equation}
[\;{\hat x}_{\mu},{\hat x}_{\nu}\;] =
i\theta_{\mu\nu}^{\rho\tau}{\hat x}_{\rho}{\hat
x}_{\tau}\;\;\;;\;\;\;\theta_{\mu\nu}^{\rho\tau} = {\rm const.}\;,
\label{noncomm2}
\end{equation}
has been  proposed  in \cite{qdef}, \cite{paolo} and \cite{lie2} at
relativistic and in \cite{daszkiewicz2} at nonrelativistic level.

Recently, there was considered the new type of quantum space -
so-called generalized quantum space-time
\begin{equation}
[\;{\hat x}_{\mu},{\hat x}_{\nu}\;] = i\theta_{\mu\nu} +
i\theta_{\mu\nu}^{\rho}{\hat x}_{\rho}\;, \label{general}
\end{equation}
which combines   canonical type with the Lie-algebraic kind  of
space-time noncommutativity. Its Hopf-algebraic realization has been
proposed in \cite{lulya}-\cite{genpogali} in the case of
relativistic symmetry and in \cite{genpogali} for its
nonrelativistic counterpart.

In this article, following the scheme proposed in \cite{krindler}
and \cite{daszrindler}, we provide the noncommutative counterparts
of Rindler space-time, associated with   generalized  twist-deformed
Poincare Hopf algebras \cite{genpogali} (see space-time
(\ref{general})). Further, we investigate the
gravito-thermodynamical radiation detected    by such generalized
twist-deformed Rindler observers in the vacuum, i.e. we find the
thermal (Hawking) spectra for twisted space-time (\ref{general}).
Particulary,  for parameter  $\theta_{\mu\nu}^{\rho}$ approaching
zero, we get the thermal spectra for canonical space-time
(\ref{wielkaslawia}) derived in  \cite{daszrindler}.

The paper is organized as follows. In  first section we recall the
basic facts concerning the generalized twist-deformed Poincare Hopf
algebras and the corresponding quantum space-times \cite{genpogali}.
The second section is devoted to the generalized  Rindler spaces,
obtained from their noncommutative Minkowski counterparts. The
deformed Hawking radiation spectra detected by  twisted Rindler
observers are derived in  section three. The final remarks are
discussed in the last section.

\section{{{Generalized twist-deformed Minkowski spaces and the corresponding Poincare Hopf structures}}}

In this section, following the paper \cite{genpogali}, we recall
basic facts related with  the generalized twist-deformed
relativistic symmetries and corresponding quantum space-times.

In accordance with the general  twist procedure
\cite{drin}-\cite{twist1}, the algebraic sectors of all discussed
below Hopf  structures remain undeformed $(\eta_{\mu\nu} =
(-,+,+,+))$
\begin{eqnarray}
&&\left[ M_{\mu \nu },M_{\rho \sigma }\right] =i\left( \eta _{\mu
\sigma }\,M_{\nu \rho }-\eta _{\nu \sigma }\,M_{\mu \rho }+\eta
_{\nu \rho }M_{\mu
\sigma }-\eta _{\mu \rho }M_{\nu \sigma }\right) \;,  \notag \\
%&~~&  \cr
&&\left[ M_{\mu \nu },P_{\rho }\right] =i\left( \eta _{\nu \rho
}\,P_{\mu }-\eta _{\mu \rho }\,P_{\nu }\right) \;\;\;,\;\;\; \left[
P_{\mu },P_{\nu }\right] =0\;,   \label{nnn}
\end{eqnarray}
while the   coproducts and antipodes  transform as follows
\begin{equation}
\Delta _{0}(a) \to \Delta _{\cdot }(a) = \mathcal{F}_{\cdot }\circ
\,\Delta _{0}(a)\,\circ \mathcal{F}_{\cdot }^{-1}\;\;\;,\;\;\;
S_{\cdot}(a) =u_{\cdot }\,S_{0}(a)\,u^{-1}_{\cdot }\;,\label{fs}
\end{equation}
where $\Delta _{0}(a) = a \otimes 1 + 1 \otimes a$, $S_0(a) = -a$
and $u_{\cdot }=\sum f_{(1)}S_0(f_{(2)})$ (we use Sweedler's
notation $\mathcal{F}_{\cdot }=\sum f_{(1)}\otimes f_{(2)}$).
Present in the above formula the twist element $\mathcal{F}_{\cdot }
\in {\mathcal U}_{\cdot}({\mathcal P}) \otimes {\mathcal
U}_{\cdot}({\mathcal P})$ satisfies the classical cocycle  condition
\begin{equation}
{\mathcal F}_{{\cdot }12} \cdot(\Delta_{0} \otimes 1) ~{\cal
F}_{\cdot } = {\mathcal F}_{{\cdot }23} \cdot(1\otimes \Delta_{0})
~{\mathcal F}_{{\cdot }}\;, \label{cocyclef}
\end{equation}
and the normalization condition
\begin{equation}
(\epsilon \otimes 1)~{\cal F}_{{\cdot }} = (1 \otimes
\epsilon)~{\cal F}_{{\cdot }} = 1\;, \label{normalizationhh}
\end{equation}
with ${\cal F}_{{\cdot }12} = {\cal F}_{{\cdot }}\otimes 1$ and
${\cal F}_{{\cdot }23} = 1 \otimes {\cal F}_{{\cdot }}$.

The corresponding to the above Hopf structure
 space-time is defined as  quantum representation space (Hopf
module) for quantum Poincare algebra, with action of the deformed
symmetry generators satisfying suitably deformed Leibnitz rules
\cite{bloch}, \cite{3b}, \cite{chi}. The action of Poincare algebra
on its Hopf module of functions depending on space-time coordinates
${x}_\mu$ is given by
\begin{equation}
P_{\mu }\rhd f(x)=i\partial _{\mu }f(x)\;\;\;,\;\;\; M_{\mu \nu
}\rhd f(x)=i\left( x_{\mu }\partial _{\nu }-x_{\nu }\partial _{\mu
}\right) f(x)\;,  \label{a1}
\end{equation}
while the $\star_{.}$-multiplication of arbitrary two functions  is
defined as follows
\begin{equation}
f({x})\star_{{\cdot}} g({x}):= \omega\circ\left(
 \mathcal{F}_{\cdot}^{-1}\rhd  f({x})\otimes g({x})\right) \;.
\label{star}
\end{equation}
In the above formula $\mathcal{F}_{\cdot}$ denotes  twist factor in
the differential representation (\ref{a1}) and $\omega\circ\left(
a\otimes b\right) = a\cdot b$.

In the article \cite{genpogali}, there have been considered three
(all possible) types  of  Abelian and generalized twist factors
($a\wedge b = a \otimes b -b \otimes a$)\footnote{Indecies $k$, $l$ are fixed and different than $i$.}:\\
 \begin{eqnarray}
   i)\;\;\;\mathcal{F}_{\theta_{kl},{\kappa}}  = \exp\;i\left[\frac{1}{2{\kappa}}P_k
\wedge M_{i0} + \theta_{kl}P_{k}\wedge P_{l}\right] \;, \label{rge1}
\end{eqnarray}
\begin{eqnarray}
  ii)\;\;\; \mathcal{F}_{\theta_{0i},\hat{\kappa}}  =
\exp \;i\left[\frac{1}{2\hat{\kappa}}P_0 \wedge M_{kl} +
{{\theta}_{0i}}P_{0 }\wedge P_{i}\right]  \;, \label{rge2}
\end{eqnarray}
and
\begin{eqnarray}
  iii)\;\;\; \mathcal{F}_{\theta_{0i},\bar{\kappa}}  =
\exp \;i\left[\frac{1}{2\bar{\kappa}}P_i \wedge M_{kl} +
{{\theta}_{0i}}P_{0 }\wedge P_{i}\right] \;, \label{rge3}
\end{eqnarray}
leading to the following generalized quantum space-times (see (\ref{general})):\\
\begin{eqnarray}
i)\;\;\;&&[\,x_0,x_a\,]_{{\star}_{\theta_{kl},{\kappa}}}
=\frac{i}{\kappa}x_i\delta_{ak}\;,\label{spacetime1} \\
&&[\,x_a,x_b\,]_{{\star}_{\theta_{kl},{\kappa}}}
=2i\theta_{kl}(\delta_{ak}\delta_{bl} - \delta_{al}\delta_{bk}) +
\frac{i}{\kappa}x_0 (\delta_{ia}\delta_{kb} -
\delta_{ka}\delta_{ib}) \;,\nonumber
\end{eqnarray}
\begin{eqnarray}
ii)\;\;\;&&[\,x_0,x_a\,]_{{\star}_{\theta_{0i},{\hat \kappa}}}
=\frac{i}{{\hat \kappa}}(\delta_{la}x_k - \delta_{ka}x_l) +
2i\theta_{0i}\delta_{ia} \;, \nonumber\\
&&[\,x_a,x_b\,]_{{\star}_{\theta_{0i},{\hat \kappa}}} =0
\;,\label{spacetime2}
\end{eqnarray}
and
\begin{eqnarray}
iii)\;\;\;&&[\,x_0,x_a\,]_{{\star}_{\theta_{0i},{\bar \kappa}}}
=2i\theta_{0i}\delta_{ia}\;,\nonumber\\
&&[\,x_a,x_b\,]_{{\star}_{\theta_{0i},{\bar \kappa}}} =
\frac{i}{{\bar \kappa}}\delta_{ib}(\delta_{ka}x_l - \delta_{la}x_k)
+ \frac{i}{{\bar \kappa}}\delta_{ia}(\delta_{lb}x_k -
\delta_{kb}x_l) \;,\label{spacetime3}
\end{eqnarray}
respectively, with  star product given by the formula (\ref{star}).

The corresponding Poincare Hopf structures have been provided in
\cite{genpogali} as well. However, due to their complicated form, in
this article, we recall as an example only one Poincare Hopf
algebra, associated with first twist factor (\ref{rge1}).  In
accordance with mentioned above twist procedure its algebraic sector
remains classical (see formula (\ref{nnn})), while the coproducts
take the form (see formula (\ref{fs}))
\begin{eqnarray}
\Delta_{\theta_{kl},{\kappa}}(P_\mu)&=&\Delta
_0(P_\mu)+\sinh(\frac{1}{2 {\kappa}} P_k )\wedge
\left(\eta_{i \mu}P_0 -\eta_{0 \mu}P_i \right)\label{coa1}\\
&+&(\cosh(\frac{1}{2 {\kappa}}  P_k )-1)\perp \left(\eta_{i \mu}P_i
-\eta_{0 \mu}P_0 \right)\;,\notag
\end{eqnarray}
\begin{eqnarray}
\Delta_{\theta_{kl},{\kappa}}(M_{\mu\nu})&=&\Delta_0(M_{\mu\nu})+\frac{1}{2
{\kappa}}M_{i 0 }\wedge  \left(\eta_{\mu
k }P_\nu-\eta_{\nu k}P_\mu\right)\nonumber\\
&+&i\left[M_{\mu\nu},M_{i 0 }\right]\wedge
\sinh(\frac{1}{2 {\kappa}} P_k ) \notag \\
&-&\left[\left[%
M_{\mu\nu},M_{i 0 }\right],M_{i 0 }\right]\perp
(\cosh(\frac{1}{2 {\kappa}}  P_k  )-1) \nonumber \\
&+&\frac{1}{2 {\kappa}}M_{i 0 }\sinh(\frac{1}{2 {\kappa}} P_k )\perp
 \left(\psi_k P_i -\chi_k P_0 \right) \label{coa100} \\
&-&\frac{1}{2 {\kappa}} \left(\psi_k P_0 -\chi_k P_i \right)\wedge
M_{i 0 }(\cosh(\frac{1}{2 {\kappa}} P_k )-1) \notag
%&~~&\nonumber\\
\end{eqnarray}
\begin{eqnarray}
&-& \theta ^{k l }[(\eta _{k \mu }P_{\nu }-\eta _{k \nu }\,P_{\mu
})\otimes P_{l }+P_{k}\otimes (\eta_{l
\mu}P_{\nu}-\eta_{l \nu}P_{\mu})]\nonumber\\
%&~~&\nonumber\\
&+& \theta_{kl}[(\eta _{l \mu }P_{\nu }-\eta _{l \nu }\,P_{\mu
})\otimes P_{k }+P_{l}\otimes (\eta_{k \mu}P_{\nu}-\eta_{k
\nu}P_{\mu})]\nonumber\\
&~~&\nonumber\\
%\end{eqnarray}
%\begin{eqnarray}
&+& \theta_{kl}\left[\left[M_{\mu\nu},M_{i 0 }\right],P_k\right]
\perp
\sinh(\frac{1}{2 {\kappa}} P_k )P_l\notag\\
&-& \theta_{kl}\left[\left[M_{\mu\nu},M_{i 0 }\right],P_l\right]
\perp \sinh(\frac{1}{2 {\kappa}} P_k )P_k
\notag\\
&+&i\theta_{kl}\left[\left[\left[%
M_{\mu\nu},M_{i 0 }\right],M_{i 0 }\right],P_k\right] \wedge
(\cosh(\frac{1}{2 {\kappa}}  P_k  )-1)P_l
\notag\\
&-& i\theta_{kl}\left[\left[\left[%
M_{\mu\nu},M_{i 0 }\right],M_{i 0 }\right],P_l\right] \wedge
(\cosh(\frac{1}{2 {\kappa}}  P_k  )-1)P_k\;,\notag
\end{eqnarray}
with   $a\perp b=a\otimes b+b\otimes a$, $\psi_\gamma =\eta_{j
\gamma }\eta_{l i}-\eta_{i \gamma }\eta_{lj}$ and $\chi_\gamma
=\eta_{j \gamma }\eta_{k i}-\eta_{i \gamma }\eta_{k j}$. The two
remaining Hopf structures corresponding to the twist factors
(\ref{rge2}) and (\ref{rge3}) look  similar to the coproducts
(\ref{coa1}) and (\ref{coa100}).

Of course, if parameter $\theta^{\mu\nu}$  goes  to zero and
parameters $\kappa$, $\hat{\kappa}$ and $\bar{\kappa}$ approach
infinity, the above space-times and corresponding  Hopf structures
become undeformed. Besides,  for fixed (different than zero)
parameters $\theta^{kl}$ and $\theta^{0i}$, and parameters $\kappa$,
$\hat\kappa$ and $\bar\kappa$ approaching infinity, we get twisted
canonical Minkowski space provided in \cite{chi}. Moreover, for
parameters $\theta^{kl}$ and $\theta^{0i}$ running to zero, and
fixed parameters $\kappa$, $\hat\kappa$ and $\bar\kappa$, we recover
the Lie-algebraically deformed relativistic spaces introduced in
\cite{lie1} and \cite{lie2}.

\section{{{Generalized twist-deformed Rindler space-times}}}

Let us now find  the   twisted Rindler spaces associated with  the
generalized  Minkowski space-times provided in pervious section. In
this aim we proceed with the algorithm proposed in \cite{krindler}
and \cite{daszrindler} for $\kappa$- and  twist-deformed Minkowski
space-times, respectively.

 We define such (Rindler) space-times as the
quantum spaces with noncommutativity given by the proper
$\ast$-multiplications. This  new $\ast$-multiplications are defined
by the new $\mathcal{Z}$-factors, which can be get from relativistic
twist
factors (\ref{rge1})-(\ref{rge3}) as follows: \\

{\bf i)} Firstly, we take the standard transformation rules from
commutative Minkowski space (described by $x_\mu$ variables) to the
accelerated and commutative as well (Rindler) space-time ($z_\mu$)
\cite{rindler}
\begin{eqnarray}
x_0 &=& z_1\, {\rm sinh} (az_0)\;,\label{strans1}\\
x_1 &=& z_1\, {\rm cosh} (az_0)\;,\label{strans2}\\
x_2 &=& z_2\;,\label{strans3} \\
x_3 &=& z_3\;,\label{strans4}
\end{eqnarray}
where $a$ denotes the acceleration parameter, i.e. we have chosen
function $N(z_1) = z_1$ in formulas (\ref{trans1})-(\ref{trans4}).\\

{\bf ii)} Further, we  rewrite  the Minkowski twist factors
(\ref{rge1})-(\ref{rge3}) (depending on commutative $x_\mu$
variables and defining the $\star$-multiplication (\ref{star})) in
terms of $z_\mu$ variables. \\

In such a way, we get three  following $\mathcal{Z}$-factors and the
corresponding Rindler
quantum spaces:\\
\\
$i)$  Rindler   space-time associated with twisted relativistic
space $i)$ (see formula (\ref{spacetime1})).

In such a case, due to the transformation  rules
(\ref{strans1})-(\ref{strans4})\footnote{One can  find, that
${\partial_{ x_0}} = (-{\rm sinh}(az_0){\partial_{ z_1}} + ({{\rm
cosh}(az_0)}/{az_1}){\partial_{ z_0}})$ and ${\partial_{ x_1}} =
({\rm cosh}(az_0){\partial_{ z_1}} - ({{\rm
sinh}(az_0)}{az_1}){\partial_{ z_0}})$.}, the proper
$\ast_{\theta_{kl},\kappa}$-product
 takes the form
\begin{equation}
f({z})\ast_{{\theta}_{kl},\kappa} g({z}):= \omega\circ\left(
 \mathcal{Z}_{\theta_{kl},\kappa}^{-1}\rhd  f({z})\otimes g({z})\right) \;,
\label{rstar}
\end{equation}
where
\begin{eqnarray}
{\mathcal{Z}}_{\theta_{kl},\kappa }^{-1} &=& \exp -i\left(
\delta_{k1} \theta_{1l}\;
f_{1}(z_0,z_1,\partial_{z_0},\partial_{z_1}) \wedge \partial_{z_l}+
\delta_{l1} \theta_{k1}\;
\partial_{z_k}\wedge f_{1}(z_0,z_1,\partial_{z_0},\partial_{z_1})+
\right.\nonumber \\
&+&\left. \delta_{k2}\delta_{l3}\theta_{23}\;\partial_{z_2}\wedge
\partial_{z_3}
 +
\delta_{k3}\delta_{l2}\theta_{32}\;\partial_{z_3}\wedge
\partial_{z_2} +
\right.
\nonumber \\
&+&\left. \frac{1}{2\kappa} \delta_{k1}
f_{1}(z_0,z_1,\partial_{z_0},\partial_{z_1})\wedge
(z_if_{0}(z_0,z_1,\partial_{z_0},\partial_{z_1}) -
g_{0}(z_0,z_1)\partial_{z_i}) +\right.\nonumber \\
&+&\left. \frac{1}{2\kappa} \delta_{i1}\;\partial_{z_k} \wedge
(g_{1}(z_0,z_1) f_{0}(z_0,z_1,\partial_{z_0},\partial_{z_1}) -
g_{0}(z_0,z_1)f_{1}(z_0,z_1,\partial_{z_0},\partial_{z_1}))
\right.\label{wielkaslawia200} \\
&+&\left. \frac{1}{2\kappa} \delta_{k2}\delta_{i3}\;\partial_{z_2}
\wedge (z_3 f_{0}(z_0,z_1,\partial_{z_0},\partial_{z_1}) -
g_{0}(z_0,z_1)\partial_{z_3})+ \right.\nonumber \\
&+&\left. \frac{1}{2\kappa} \delta_{k3}\delta_{i2}\;\partial_{z_3}
\wedge (z_2 f_{0}(z_0,z_1,\partial_{z_0},\partial_{z_1}) -
g_{0}(z_0,z_1)\partial_{z_2})
\right)=\nonumber \\
&=& \exp \left({\cal A}_{\theta_{kl},\kappa}(z,\partial_z) \wedge
{\cal B}_{\theta_{kl},\kappa}(z,\partial_z) \right)= \exp {\cal
O}_{\theta_{kl},\kappa}(z,\partial_z)\;, \nonumber
\end{eqnarray}
and
\begin{eqnarray}
f_{0}(z_0,z_1,\partial_{z_0},\partial_{z_1}) &=& -{\rm
sinh}(az_0)i{\partial_{ z_1}} + ({{\rm
cosh}(az_0)}/{az_1})i{\partial_{ z_0}}\;,\label{funkcja0}\\
f_{1}(z_0,z_1,\partial_{z_0},\partial_{z_1}) &=& {\rm
cosh}(az_0)i{\partial_{ z_1}} - ({{\rm
sinh}(az_0)}/{az_1})i{\partial_{ z_0}}\;,\label{funkcja1}\\
g_{0}(z_0,z_1) &=&z_1\, {\rm sinh} (az_0) \;\;\;,\;\;\;
g_{1}(z_0,z_1) =z_1\, {\rm cosh} (az_0) \;.\label{gfunkcje}
\end{eqnarray}
However, to simplify, we consider the following differential
operator
\begin{equation}
({ {\mathcal{Z}}}_{\theta_{kl},\kappa}^{\rm Linear})^{-1} = 1+{\cal
O}_{\theta_{kl},\kappa}(z,\partial_z)\;,\label{linear}
\end{equation}
which  contains only the  terms linear in deformation parameter
$\theta^{kl}$ and $\kappa$\footnote{We
 look only  for the corrections   to
Hawking radiation   linear in deformation parameter.}. Hence, the
linearized ${\hat \ast}$-Rindler multiplication is given by the
formula (\ref{rstar}), but with differential operator (\ref{linear})
instead the complete  factor ${\mathcal{Z}}_{\theta_{kl},\kappa
}^{-1}$. Consequently, for $f(z) = z_\mu$ and $g(z) = z_\nu$, we get
\begin{equation}
[\;{ z}_{\mu},{ z}_{\nu}\;]_{{\hat \ast}_{{\theta}_{kl}},\kappa} =
\left[({\cal A}_{\theta_{kl},\kappa}(z,\partial_z) z_\mu)({\cal
B}_{\theta_{kl},\kappa}(z,\partial_z)z_\nu) -({\cal
B}_{\theta_{kl},\kappa}(z,\partial_z)z_\mu)({\cal
A}_{\theta_{kl},\kappa}(z,\partial_z)z_\nu) \right]
\label{canonrindler}
\end{equation}
with $f_2(z,\partial_z) = i\partial_{z_2}$, $f_3(z,\partial_z)=
i\partial_{z_3}$. The above commutation relations define the
generalized twist-deformed Rindler space-time associated with
generalized Minkowski space
(\ref{spacetime1}). \\
\\
$ii)$ Rindler   space-time associated with generalized
twist-deformed Minkowski  space $ii)$ (see (\ref{spacetime2})).

Here, due to the rules (\ref{strans1})-(\ref{strans4}), the
$\ast_{\theta_{0i},\hat\kappa}$-multiplication look as  follows
\begin{equation}
f({z})\ast_{\theta_{0i},\hat\kappa} g({z})= \omega\circ\left(
 \mathcal{Z}_{\theta_{0i},\hat\kappa}^{-1}\rhd  f({z})\otimes g({z})\right) \;,
\label{rstar2}
\end{equation}
where
\begin{eqnarray}
{\mathcal{Z}}_{\theta_{0i},\hat\kappa }^{-1} &=& \exp -i\left(
\delta_{i1}\theta_{01}\;
f_{0}(z_0,z_1,\partial_{z_0},\partial_{z_1}) \wedge
f_{1}(z_0,z_1,\partial_{z_0},\partial_{z_1}) +
\right.\nonumber\\
&+& \left. \delta_{i2}\theta_{02}\;
f_{0}(z_0,z_1,\partial_{z_0},\partial_{z_1}) \wedge \partial_{z_2} +
\delta_{i3}\theta_{03}\;
f_{0}(z_0,z_1,\partial_{z_0},\partial_{z_1}) \wedge \partial_{z_3} +
\right.\nonumber \\
&+&\left.
\frac{1}{2\hat{\kappa}}\delta_{k2}\delta_{l3}\;f_{0}(z_0,z_1,\partial_{z_0},\partial_{z_1})
\wedge (z_2 \partial_{z_3} - z_3\partial_{z_2}) +\right.
\nonumber\\
&+&\left.
\frac{1}{2\hat{\kappa}}\delta_{k3}\delta_{l2}\;f_{0}(z_0,z_1,\partial_{z_0},\partial_{z_1})
\wedge (z_3 \partial_{z_2} - z_2\partial_{z_3}) +\right.
\label{wielkaslawia201}\\
&+&\left.
\frac{1}{2\hat{\kappa}}\delta_{k1}\;f_{0}(z_0,z_1,\partial_{z_0},\partial_{z_1})
\wedge (g_{1}(z_0,z_1) \partial_{z_l} -
z_lf_{1}(z_0,z_1,\partial_{z_0},\partial_{z_1}))+\right.
\nonumber\\
&+&\left.
\frac{1}{2\hat{\kappa}}\delta_{l1}\;f_{0}(z_0,z_1,\partial_{z_0},\partial_{z_1})
\wedge (z_k f_{1}(z_0,z_1,\partial_{z_0},\partial_{z_1})-
g_{1}(z_0,z_1) \partial_{z_k})\right)=
\nonumber\\
&=&  \exp \left({\cal C}_{\theta_{0i},\hat\kappa}(z,\partial_z)
\wedge {\cal D}_{\theta_{0i},\hat\kappa}(z,\partial_z) \right) =
\exp {\cal O}_{\theta_{0i},\hat\kappa}(z,\partial_z)\;,\nonumber
\end{eqnarray}
Consequently, the corresponding (linearized) Rindler space-time
takes the form
\begin{equation}
[\;{ z}_{\mu},{ z}_{\nu}\;]_{{\hat \ast}_{\theta_{0i},\hat\kappa}} =
\left[({\cal C}_{{\theta_{0i},\hat\kappa}}(z,\partial_z)
z_\mu)({\cal D}_{{\theta_{0i},\hat\kappa}}(z,\partial_z)z_\nu)
-({\cal D}_{{\theta_{0i},\hat\kappa}}(z,\partial_z)z_\mu)({\cal
C}_{{\theta_{0i},\hat\kappa}}(z,\partial_z)z_\nu) \right]
\label{lierindler}
\end{equation}
where we use  the linearized approximation to
${{\mathcal{Z}}}_{\theta_{0i},\hat\kappa }^{-1}$ (see (\ref{linear})).\\
\\
$iii)$  Rindler space-time associated with relativistic
twist-deformed space $iii)$ (see formula (\ref{spacetime3})).

In such a  case, the $\ast_{\theta_{0i},\bar\kappa}$-multiplication
takes the form
\begin{equation}
f({z})\ast_{\theta_{0i},\bar\kappa} g({z})= \omega\circ\left(
 \mathcal{Z}_{\theta_{0i},\bar\kappa}^{-1}\rhd  f({z})\otimes g({z})\right) \;,
\label{rstar3}
\end{equation}
with factor
\begin{eqnarray}
{\mathcal{Z}}_{\theta_{0i},\bar\kappa }^{-1}&=& \exp
-i\left(\delta_{i1}\theta_{01}\;
f_{0}(z_0,z_1,\partial_{z_0},\partial_{z_1}) \wedge
f_{1}(z_0,z_1,\partial_{z_0},\partial_{z_1}) +
\right.\nonumber\\
&+& \left. \delta_{i2}\theta_{02}\;
f_{0}(z_0,z_1,\partial_{z_0},\partial_{z_1}) \wedge \partial_{z_2} +
\delta_{i3}\theta_{03}\;
f_{0}(z_0,z_1,\partial_{z_0},\partial_{z_1}) \wedge \partial_{z_3} +
\right.\nonumber \\
&+&\left. \frac{1}{2\bar\kappa}
\delta_{i1}\;f_{1}(z_0,z_1,\partial_{z_0},\partial_{z_1}) \wedge
(z_k
\partial_{z_l}- z_l\partial_{z_k})+
\right. \label{twistkwad} \\
&+&\left. \frac{1}{2\bar\kappa} \delta_{k1}\;\partial_{z_i} \wedge
(g_{1}(z_0,z_1)\partial_{z_l} -
z_lf_{1}(z_0,z_1,\partial_{z_0},\partial_{z_1})) +
\right.\nonumber \\
&+&\left. \frac{1}{2\bar\kappa} \delta_{l1}\;\partial_{z_i} \wedge
(z_kf_{1}(z_0,z_1,\partial_{z_0},\partial_{z_1}) -
g_{1}(z_0,z_1)\partial_{z_k})\right) =
\nonumber \\
&=& \exp \left({\cal E}_{\theta_{0i},\bar\kappa}(z,\partial_z)
\wedge {\cal G}_{\theta_{0i},\bar\kappa}(z,\partial_z) \right)
 = \exp {\cal
O}_{\theta_{0i},\bar\kappa}(z,\partial_z)\;.\nonumber
\end{eqnarray}
Then,  the (linearized)  Rindler space looks as follows
\begin{equation}
[\;{ z}_{\mu},{ z}_{\nu}\;]_{{\hat \ast}_{\theta_{0i},\bar\kappa}} =
\left[({\cal E}_{{\theta_{0i},\bar\kappa}}(z,\partial_z)
z_\mu)({\cal G}_{{\theta_{0i},\bar\kappa}}(z,\partial_z)z_\nu)
-({\cal G}_{{\theta_{0i},\bar\kappa}}(z,\partial_z)z_\mu)({\cal
E}_{{\theta_{0i},\bar\kappa}}(z,\partial_z)z_\nu) \right]
\label{quarindler}
\end{equation}
with ${{\hat \ast}_{\theta_{0i},\bar\kappa}}$-multiplication defined
by the linear
approximation to (\ref{twistkwad}) (see (\ref{linear})).\\

Obviously, for both deformation parameters $\theta^{kl}$ and
$\theta^{0i}$ approaching zero, and all parameters $\kappa$,
$\hat\kappa$ and $\bar\kappa$ running to infinity, the above
generalized Rindler space-times become classical. It should be also
noted, that for fixed (different than zero) parameters $\theta^{kl}$
and $\theta^{0i}$, and parameters $\kappa$, $\hat\kappa$ and
$\bar\kappa$ approaching infinity, we get twisted canonical  Rindler
space provided in \cite{daszrindler}. On the other side, for
parameters $\theta^{kl}$ and $\theta^{0i}$ running to zero, and
fixed parameters $\kappa$, $\hat\kappa$ and $\bar\kappa$, we recover
the Lie-algebraically deformed Rindler spaces introduced in
\cite{daszrindler} as well.

\section{{{Hawking thermal spectra  for  generalized twist-defo-rmed  Rindler space-times}}}

In this   section we find the corrections to the
gravito-thermodynamical process, which occur  in generalized
twist-deformed  space-times.

As it was mentioned in Introduction, such effects as Hawking
radiation \cite{hawking1},  can be observed in vacuum by uniformly
accelerated observer \cite{devies}, \cite{unruh}. First of all,
following \cite{krindler} and \cite{daszrindler}, we  recall the
calculations performed for
 gravito-thermodynamical process in commutative relativistic space-time
\cite{simpl1}, \cite{simpl2}. Firstly, we consider the on-shell
plane wave corresponding to the massless mode with positive
frequency $\hat{\omega}$ moving in $x_1=x$ direction of Minkowski
space ($x_0 =t$)
\begin{equation}
\phi (x,t) = \exp \left(\hat{\omega}x- \hat{\omega}t \right) \;.
\label{field1}
\end{equation}
In terms of Rindler variables this plane wave takes the form
$(z_0=\tau, z_1=z)$
\begin{equation}
\phi (x(z,\tau),t(z,\tau)) \equiv \phi (z,\tau)  = \exp
\left(i\hat{\omega}z {\rm e}^{-a\tau} \right) \;, \label{field2}
\end{equation}
i.e. it  becomes nonmonochromatic  and instead has the frequency
spectrum $f(\omega)$, given by Fourier transform
\begin{equation}
 \phi (z,\tau) = \int_{-\infty}^{+\infty}\frac{d\omega}{2\pi}
 f(\omega){\rm
 e}^{-i\omega \tau}
\;.  \label{trans}
\end{equation}
The corresponding power spectrum is given by $P(\omega) =
|f(\omega)|^2$ and the function $f(\omega)$ can be obtained  by
inverse Fourier transform
\begin{equation}
f(\omega)  =\int_{-\infty}^{+\infty}d\tau\,{\rm e}^{i\hat{\omega}z
{\rm e}^{-a\tau}}{\rm
 e}^{i\omega \tau} = \left(-\frac{1}{a}\right)\left({\hat \omega}z\right)^{i\omega/a}\Gamma
 \left(-\frac{i\omega}{a}\right){\rm e}^{\pi\omega/2a}
\;,  \label{invtrans}
\end{equation}
where $\Gamma (x)$ denotes the gamma function \cite{gamma}. Then,
since
\begin{equation}
\left|\Gamma\left(\frac{i{ \omega}}{a}\right)\right|^2  =
\frac{\pi}{({ \omega}/a){\rm sinh}(\pi{ \omega}/a)} \;,
\label{gamma}
\end{equation}
we get the following power spectrum at negative frequency
\begin{equation}
\omega P(-\omega) = \omega |f(-\omega)|^2  = \frac{2\pi/a}{{\rm
e}^{2\pi\omega/a}-1}\;, \label{power}
\end{equation}
which corresponds to the  Planck factor $\left({\rm
e}^{{\hbar\omega}/kT} -1 \right)$ associated with temperature $T =
\hbar a/2\pi kc$ (the temperature of radiation seen by Rindler
observer (see formula (\ref{tempe1}))).

Let us now turn to the case of generalized twist-deformed
space-times provided in pervious section. In order to find the power
spectra for such deformed Rindler spaces, we start with the
(fundamental) formula (\ref{field2}) for scalar field, equipped with
the twisted (linearized) ${\hat \ast}$-multiplications
\begin{equation}
 \phi^{{\rm Twisted}}_{\cdot,\cdot} (z,\tau)  = \exp
\left(i\hat{\omega}z{\hat \ast}_{\cdot,\cdot} {\rm e}^{-a\tau}
\right) \;. \label{field222}
\end{equation}
Then
\begin{equation}
f^{{\rm Twisted}}_{\cdot,\cdot}(\omega)  =
\int_{-\infty}^{+\infty}d\tau\,{\rm e}^{i\hat{\omega}z{\hat
\ast}_{\cdot,\cdot} {\rm e}^{-a\tau}} {\hat \ast}_{\cdot,\cdot} {\rm
 e}^{i\omega \tau}
\;, \label{modyfication}
\end{equation}
and, in accordance with the pervious  considerations, we
get\footnote{We only take under consideration  the terms linear in
deformation parameters $\theta^{kl}$, $\theta^{0i}$, $\kappa$,
$\hat\kappa$ and $\bar\kappa$.}
\begin{eqnarray}
f^{{\rm Twisted}}_{\cdot,\cdot}(\omega)  &=& f(\omega) +
\int_{-\infty}^{+\infty}d\tau\, \omega\circ\left(
 {\cal
O}_{\cdot,\cdot}(\tau,z,\partial_\tau,\partial_z)\rhd  {\rm
e}^{i\hat{\omega}z {\rm e}^{-a\tau}}\otimes {\rm
 e}^{i\omega \tau}\right) +   \label{modyfication1}\\
 &+&
\int_{-\infty}^{+\infty}d\tau\, {\rm e}^{i\hat{\omega}z {\rm
e}^{-a\tau}} {\rm
 e}^{i\omega \tau}\omega\circ\left(
 {\cal
O}_{\cdot,\cdot}(\tau,z,\partial_\tau,\partial_z)\rhd i{\hat \omega}
z \otimes {\rm e}^{-a\tau} \right)\;; \nonumber
\end{eqnarray}
 the corresponding (twisted) power spectrum is defined as
\begin{eqnarray}
\omega P^{{\rm Twisted}}_{\cdot,\cdot}(-\omega) &=&
\omega\left|f^{{\rm Twisted}}_{\cdot,\cdot}(-\omega) \right|^2\;.
\label{modyfication2}
\end{eqnarray}
Consequently, due to the form of linearized Rindler factors
${{\mathcal{Z}}}_{\cdot,\cdot }^{\rm Linear}$, we have:\\

$i)$ The thermal spectrum for generalized Rindler space $i)$.

 In such a case the operator ${\cal
O}_{\cdot,\cdot}(\tau,z,\partial_\tau,\partial_z) = {\cal
O}_{\theta_{kl},\kappa}(\tau,z,\partial_\tau,\partial_z)$ is given
by the formula (\ref{wielkaslawia200}) and
\begin{eqnarray}
f^{{\rm Twisted}}_{\theta_{kl},\kappa}(\omega) &=& f(\omega)+
\frac{i\delta_{k1}\;z_i\;\omega\;{\hat \omega}}{2\kappa
az}\int_{-\infty}^{+\infty}d\tau\,{\rm e}^{i\hat{\omega}z {\rm
e}^{-a\tau}} {\rm
 e}^{i\omega \tau}{\rm e}^{-a\tau}+\;\;\;\;\;\;\;\;\;\;\;\;\nonumber\\
&&\;\;\;\;\;\;\;\;\;\;\;\;\;\;\;\;\;\;\;\;\;-\;\frac{\delta_{k1}\;z_i\;{\hat
\omega}}{2\kappa z}\int_{-\infty}^{+\infty}d\tau\,{\rm
e}^{i\hat{\omega}z {\rm e}^{-a\tau}} {\rm
 e}^{i\omega \tau}{\rm e}^{-a\tau}
 \;.
\label{modyficationp}
\end{eqnarray}
The above integral can be   evaluated with  help of standard
identity for Gamma function
\begin{equation}
\Gamma(y+1) = y\Gamma (y) \label{gammma}\;;
\end{equation}
one gets
\begin{equation}
f^{{\rm Twisted}}_{\theta_{kl},\kappa}(\omega)  =
\left(-\frac{1}{a}\right)\left({\hat
\omega}z\right)^{i\omega/a}\Gamma
 \left(-\frac{i\omega}{a}\right){\rm e}^{\pi\omega/2a}\left( 1 +
 \frac{\delta_{k1}\;z_i\;\omega}{2\kappa az^2}\left[\frac{i\omega}{a}
-1\right] \right) \;. \label{calka}
\end{equation}
Then, one can check that the thermal spectrum takes the form
\begin{eqnarray}
\omega P^{{\rm Twisted}}_{\theta_{kl},\kappa}(-\omega) =
\frac{1}{T}\frac{1}{{\rm e}^{\omega/T}-1}\left( 1 -
\frac{\delta_{k1}\;z_i\; \omega}{2\kappa\pi T z^2} \right) + {\cal
O}(\kappa^2) \;, \label{calka1}
\end{eqnarray}
with Hawking temperature $T=a/2\pi$ associated with the acceleration
of generalized twist-deformed  Rindler observer $i)$. \\

$ii)$ The Hawking radiation spectrum associated with twist
deformation $ii)$.

Then, ${\cal O}_{\cdot,\cdot}(\tau,z,\partial_\tau,\partial_z) =
{\cal O}_{\theta_{0i},\hat\kappa}(\tau,z,\partial_\tau,\partial_z)$
(see formula (\ref{wielkaslawia201})) and the function \\
$f^{{\rm
Twisted}}_{\cdot,\cdot}(\omega)$ looks as follows
\begin{eqnarray}
f^{{\rm Twisted}}_{\theta_{0i},\hat\kappa}(\omega) &=& f(\omega)+
i\left(\theta_{01} + \frac{1}{2\hat\kappa}(\delta_{l1}z_k-
\delta_{k1}z_l)\right)\frac{\omega{\hat
\omega}}{az}\int_{-\infty}^{+\infty}d\tau\,{\rm e}^{i\hat{\omega}z
{\rm e}^{-a\tau}} {\rm
 e}^{i\omega \tau}{\rm e}^{-a\tau}+\nonumber\\
&-&\left(\theta_{01} + \frac{1}{2\hat\kappa}(\delta_{l1}z_k-
\delta_{k1}z_l)\right)\frac{{\hat
\omega}}{z}\int_{-\infty}^{+\infty}d\tau\,{\rm e}^{i\hat{\omega}z
{\rm e}^{-a\tau}} {\rm
 e}^{i\omega \tau}{\rm e}^{-a\tau}
 \;.
\label{modyficationp2}
\end{eqnarray}
One can perform the above integral with respect time $\tau$
\begin{eqnarray}
f^{{\rm Twisted}}_{\theta_{0i},\hat\kappa}(\omega) &=&
\left(-\frac{1}{a}\right)\left({\hat
\omega}z\right)^{i\omega/a}\Gamma
 \left(-\frac{i\omega}{a}\right){\rm
 e}^{\pi\omega/2a}\cdot\;\;\;\;\;\;\;\;\;\;\;\;
 \nonumber\\
&&\;\;\;\;\cdot \left( 1 + \left(\theta_{01} +
\frac{1}{2\hat\kappa}(\delta_{l1}z_k-
\delta_{k1}z_l)\right)\frac{\omega}{az^2}\left[\frac{i\omega}{a}
-1\right] \right) \;,\label{calka2}
\end{eqnarray}
and, consequently, the power spectrum takes the form
\begin{eqnarray}
\omega P^{{\rm Twisted}}_{\theta_{0i},\hat\kappa}(-\omega)
&=&\frac{1}{T}\frac{1}{{\rm e}^{\omega/T}-1}\left( 1 -
\left(\theta_{01} + \frac{1}{2\hat\kappa}(\delta_{l1}z_k-
\delta_{k1}z_l)\right)\frac{ \omega}{\pi T z^2} \right)+\nonumber\\
&+& {\cal O}(\theta_{01}^2, \hat\kappa^2)\;, \label{calka12}
\end{eqnarray}
with Hawking temperature $T=a/2\pi$. \\

$iii)$ The thermal spectrum corresponding to the generalized
deformation of Rindler space $iii)$.

In the last case ${\cal
O}_{\cdot,\cdot}(\tau,z,\partial_\tau,\partial_z) = {\cal
O}_{\theta_{0i},\bar\kappa}(\tau,z,\partial_\tau,\partial_z)$ (see
formula (\ref{twistkwad})) and, we have
\begin{eqnarray}
f^{{\rm Twisted}}_{\theta_{0i},\bar\kappa}(\omega) &=& f(\omega)+
\frac{i\delta_{i1}\;\theta_{01}\;\omega\;{\hat
\omega}}{az}\int_{-\infty}^{+\infty}d\tau\,{\rm e}^{i\hat{\omega}z
{\rm e}^{-a\tau}} {\rm
 e}^{i\omega \tau}{\rm e}^{-a\tau}
+\;\;\;\;\;\;\;\;\;\;\;\;\nonumber\\
&&\;\;\;\;\;\;\;\;\;\;\;\;\;\;\;\;\;\;\;\;\;-\;
\frac{\delta_{i1}\;\theta_{01}\;{\hat
\omega}}{z}\int_{-\infty}^{+\infty}d\tau\,{\rm e}^{i\hat{\omega}z
{\rm e}^{-a\tau}} {\rm
 e}^{i\omega \tau}{\rm e}^{-a\tau}
 \;.
\label{modyficationp3}
\end{eqnarray}
After integration (see identity for Gamma function (\ref{gammma})),
one gets
\begin{equation}
f^{{\rm Twisted}}_{\theta_{0i},\bar\kappa}(\omega)  =
\left(-\frac{1}{a}\right)\left({\hat
\omega}z\right)^{i\omega/a}\Gamma
 \left(-\frac{i\omega}{a}\right){\rm e}^{\pi\omega/2a}\left( 1 +
 \frac{\delta_{i1}\;\theta_{01}\;\omega}{ az^2}\left[\frac{i\omega}{a}
-1\right] \right) \;,
 \label{calka3}
\end{equation}
and
\begin{eqnarray}
\omega P^{{\rm Twisted}}_{\theta_{0i},\bar\kappa}(-\omega) =
\frac{1}{T}\frac{1}{{\rm e}^{\omega/T}-1}\left( 1 -
\frac{\delta_{i1}\;\theta_{01}\; \omega}{\pi T z^2} \right) + {\cal
O}(\theta_{01}^2) \;, \label{calka13}
\end{eqnarray}
respectively.\\

Of course, for  deformation parameter $\theta_{01}$ approaching
zero, and  parameters $\kappa$ and $\hat\kappa$ running to infinity,
the above corrections disappear. Besides, one can observe, that for
deformation parameters $\kappa$ and $\hat\kappa$ going to infinity,
we get the thermal spectrum for canonical Rindler space-time
provided in \cite{daszrindler}.

\section{{{Final remarks}}}
In this article we provide  linear version of three generalized
twist-deformed Rindler spaces. All of them  correspond  to the
generalized twist-deformed Minkowski space-times derived   in
\cite{genpogali}. Further, we demonstrated that  there  appear
corrections to the Hawking thermal radiation, which are linear in
deformation parameters $\theta^{01}$, $\kappa$ and $\hat\kappa$.

It should be noted, that the above results can be extended in
different ways. First of all, for example, one can find the complete
form of (generalized) Rindler space-times  with the  use of complete
twist differential operators
\begin{equation}
{\cal Z}_{\cdot, \cdot}^{-1} = \exp {\cal O}_{\cdot,
\cdot}(z,\partial_z)\;, \label{complete}
\end{equation}
which appear respectively in the  formulas (\ref{rstar}),
(\ref{rstar2}) and (\ref{rstar3}). However, due to the complicated
form of  operators ${\cal O}_{\cdot, \cdot}(z,\partial_z)$  such a
problem seems to be  quite difficult to solve from technical point
of view. Besides, following the paper \cite{simpl1} (the case of
commutative Rindler space), one can find additional physical
applications for such deformed generalized Rindler space-times. The
studies in these directions already started and are  in progress.

\section*{Acknowledgments}
The author would like to thank  J. Lukierski
for valuable discussions.\\
This paper has been financially supported by Polish Ministry of
Science and Higher Education grant NN202318534.

\end{document}